\pdfoutput=1

\documentclass[11pt,twoside,a4paper,cmspaper,final,collab]{cms-tdr}
\def\svnVersion{191427}\def\cmsCernNo{2012-358}\def\cmsCernDate{\today}\def\cmsMessage{Published in the Journal of High Energy Physics as \href{http://dx.doi.org/10.1007/JHEP05(2013)065}{\doi{10.1007/JHEP05(2013)065.}}}

\begin{document}\cmsNoteHeader{TOP-11-007}

\hyphenation{had-ron-i-za-tion}
\hyphenation{cal-or-i-me-ter}
\hyphenation{de-vices}
\RCS$Revision: 180309 $
\RCS$HeadURL: svn+ssh://svn.cern.ch/reps/tdr2/papers/TOP-11-007/trunk/TOP-11-007.tex $
\RCS$Id: TOP-11-007.tex 180309 2013-04-09 14:32:39Z stadie $
\cmsNoteHeader{TOP-11-007} 
\title{Measurement of the \ttbar production cross section in the all-jet final state in pp collisions at $\sqrt{s}=7$\TeV}

\date{\today}

\abstract{
A measurement is presented of the $\ttbar$ cross section ($\sigma_{\ttbar}$) in proton-proton collisions at a centre-of-mass energy of 7\TeV, in the all-jet final state that contains at least six jets, two of which are tagged as originating from b quarks. The data correspond to an integrated luminosity of 3.54\fbinv, collected with the CMS detector at the LHC. The cross section is determined through an unbinned maximum likelihood fit of background and $\ttbar$ signal to the reconstructed mass spectrum of $\ttbar$ candidates in the data, in which events are subjected to a kinematic fit assuming a $\ttbar \to \PWp\cPqb\PWm\cPaqb\to 6$ jets hypothesis. The measurement yields $\sigma_{\ttbar} = 139 \pm 10\stat \pm 26\syst \pm 3\lum\unit{pb}$, a result consistent with those obtained in other \ttbar decay channels, as well as with predictions of the standard model.}

\hypersetup{%
pdfauthor={CMS Collaboration},%
pdftitle={Measurement of the ttbar production cross section in the all-jet final state in pp collisions at sqrt(s) = 7 TeV},%
pdfsubject={CMS},%
pdfkeywords={CMS, physics, top, cross section}}

\maketitle 

\section{Introduction}
Precise measurements of the top-quark pair (\ttbar) production cross section ($\sigma_{\ttbar}$), especially in different final states, in proton-proton (pp) collisions at the Large Hadron Collider (LHC) provide important checks of perturbative quantum chromodynamics (QCD) and therefore of the standard model (SM). Such studies are also of value in estimating backgrounds in searches for new physics. First measurements from pp collisions in 2010 at a centre-of-mass energy of $\sqrt{s}=7$\TeV, based on an integrated luminosity of $3\pbinv$, were reported by the CMS~\cite{Khachatryan:2010ez} and ATLAS~\cite{Aad:2010ey} experiments. Subsequent measurements, based on all the data collected in 2010, corresponding to an integrated luminosity of approximately 36\pbinv were performed in the dilepton ($\ttbar \to \PWp\cPqb\PWm\cPaqb \to \ell^{+}\nu_{\ell} \cPqb \ell^{-} \Pagn_{\ell} \cPaqb$)~\cite{top-11-002,Aad:2011yb} and in the lepton+jets channels ($\PWp\cPqb\PWm\cPaqb \to \ell^{+}\nu_{\ell} \cPqb\cPaq\cPq'\cPaqb$ $+$ charge conjugate states), both with~\cite{top-10-003,Aad:2012qf} and without~\cite{Aad:2012qf,top-10-002} the use of b tagging. Measurements using data collected in 2011, corresponding to integrated luminosities between 0.7 and $2.3\fbinv$, were also reported in lepton+jets and dilepton channels for contributions from $\tau\to\text{hadrons}+\nu_{\tau}$ decays~\cite{ATLAS:tau-jets,top-11-006,ATLAS:2012xh}, as well as from final states containing just electrons or muons~\cite{ATLAS:2012aa,top-11-005,top-11-003}.

This Letter presents the first measurement of \ttbar production cross section in proton-proton collisions at $\sqrt{s}=7$\TeV in the all-jet decay channel ($\PWp \cPqb \PWm \cPaqb\to\cPaq \cPq' \cPqb \cPq'' \cPaq''' \cPaqb$) by the CMS collaboration. The measurement is complementary to the previous measurements of $\sigma_{\ttbar}$, and is therefore interesting in its own right. The all-jet final state has a far larger yield of \ttbar events than the dilepton or lepton+jets channels. Moreover it does not suffer from the presence of neutrinos of large transverse momentum \pt that escape detection. However, with only jets in the final state, this channel is dominated by background from generic multijet (MJ) production. The main analysis is based on a reconstruction of the candidate events through a kinematic fit to the \ttbar hypothesis. An alternative measurement is performed using a neural-network-based selection and a different model for background.
\section{The CMS Detector}
\label{sec:detector}
The central feature of the CMS apparatus is a superconducting solenoid, 13\unit{m} in length and 6\unit{m} in diameter, which provides an axial magnetic field of 3.8\unit{T}. The bore of the solenoid is outfitted with various particle detection systems. Charged-particle trajectories are measured with silicon pixel and strip trackers, covering $0 < \phi < 2\pi$ in azimuth and $|\eta |<2.5$ in pseudorapidity, where $\eta$ is defined as $-\ln\left(\tan{\theta/2}\right)$, with $\theta$ being the polar angle of the trajectory of the particle with respect to the anticlockwise beam direction. A lead tungstate crystal electromagnetic calorimeter and a brass/scintillator hadronic calorimeter surround the tracking volume. The calorimetry provides excellent resolution in energy for electrons and jets of hadrons within $|\eta|<3.0$. Muons are measured up to $|\eta|<2.4$ using gas-ionisation detectors embedded in the steel return yoke outside the solenoid. The detector is nearly hermetic, providing accurate measurements of any imbalance in momentum in the plane transverse to the beam direction. A two-level trigger system selects the pp final states pertinent to this analysis. A detailed description of the CMS detector is available in Ref.~\cite{CMS:2008zzk}.
\section{Data and Simulation}
\label{sec:datasets}
The data correspond to an integrated luminosity of 3.54\fbinv collected in 2011. Two multijet triggers are used in this analysis. The first requires the presence of at least four jets in the calorimeter, each with $\pt > 50$\GeVc, and a fifth jet with $\pt > 40$\GeVc. The second trigger, intended for coping with the higher instantaneous luminosities in the latter stages of the data taking period, requires an additional sixth jet with $\pt > 30$\GeVc. Data taken beyond the initial 3.54\fbinv in 2011 are not used in this analysis because a more-restrictive trigger implemented for highest luminosities greatly reduced the gain in the number of selected \ttbar events. The efficiencies for the two triggers are determined from events that pass a prescaled trigger with a less-restrictive requirement of at least four jets with $\pt > 40$\GeVc that is highly efficient for \ttbar signal, as estimated from \ttbar Monte Carlo (MC) simulation. Trigger efficiencies are defined by the number of such events that pass the two tighter trigger requirements for the offline selection criteria described in Section~\ref{sec:eventselection}, relative to the number that pass just the offline selections. The combined trigger efficiency for \ttbar signal for data is 96$^{+4}_{-5}$\%, and 99.7$^{+0.3}_{-0.4}$\% for the MC simulation described below, where the uncertainties are purely statistical for the MC, and mainly systematic for the data (see Section~\ref{sec:systematics}).

The simulated \ttbar events are generated considering QCD matrix elements with up to three additional final-state partons, using \MADGRAPH~v5.1.1.0~\cite{Alwall:2011} interfaced to the \PYTHIA v6.424 MC generator~\cite{Sjostrand:2006za} for providing perturbative quantum-chromodynamic parton showering. The value of the top-quark mass in the MC is $m_\cPqt = 172.5$\GeVcc, and the proton structure is described by the parton distribution functions (PDF) CTEQ6L1~\cite{bib:cteq}. The Z2 tune~\cite{tuneZ2} is used to characterise the underlying event. The simulation includes the effects of the presence of additional, overlapping minimum-bias pp interactions (pileup) and a weighting procedure ensures that the pileup profile in the simulated events matches the one inferred from data with an average number of eight additional pileup events. Multijet events are simulated using the leading-order (LO) QCD \PYTHIA MC, and are used to check the validity of the method of analysis. However, the background from such generic multijet events is estimated from data through events in the sidebands of the signal to represent better the multijet component within the signal-dominated region.
\section{Event Selection}
\label{sec:eventselection}
Offline collections of reconstructed particles are produced using a particle-flow (PF) algorithm \cite{PFT-09-001,PFT-10-002}. Jets are clustered together from the PF particles using the anti-\kt jet algorithm~\cite{antikt}, implemented through {\textsc{FastJet}}~v2.4.2~\cite{fastjet} with a distance parameter of $R=0.5$. The PF particles can be charged or neutral hadrons, electrons, photons, or muons. By combining information from all subdetectors, the PF technique reduces considerably the size of energy corrections required for otherwise reconstructed jets. Another advantage of this technique is that it also reduces the impact of event pileup at large luminosities by discarding contributions from charged particles associated with other than the primary and secondary vertices of the most energetic pp interaction, defined by the largest value of the sum of $\sum \pt^2$ over all the associated tracks. Effects from pileup, including that of energy deposition from neutral hadrons, are reduced further by using the \textsc{FastJet} pileup subtraction procedure~\citep{Cacciari:2007fd,Cacciari:2008gn}. To minimise contamination from jet candidates generated through electronics noise or from electrons reconstructed as jets, evidence must be present of energy deposition in the calorimeter that can be attributed to charged hadrons originating from the jet, and no more than 99\% of the jet energy can be attributed to a combination of photons, electrons, and neutral hadrons.

Since this analysis focuses on purely hadronic decays of the \ttbar system, with each top quark producing a minimum of three jets ($\cPqt\to\cPqb\PW \to \cPqb\cPaq\cPq'$) in the final state, candidate events are required to have at least four jets with $\pt > 60 \GeVc$, a fifth jet with $\pt > 50 \GeVc$, and a sixth jet with $\pt > 40 \GeVc$. All jets must be observed within $|\eta| < 2.4$, which corresponds to a fiducial region in the inner tracker acceptance. Additional jets are considered for use in the kinematic fit described later (Section~\ref{sec:kinfit}) if they have $\pt > 30 \GeVc$ and $|\eta| < 2.4$.

Jets originating from bottom quarks are classified as b-tagged jets through an algorithm based on the reconstruction of secondary vertices~\cite{CMS-PAPER-BTV-11-004}. Only secondary vertices with at least three tracks are considered for designation as b jets. A discriminant $d_{\textrm{B}}$ reflects the significance of the decay length (the distance between the primary and secondary vertices) in each event. A restrictive working point chosen for this algorithm, with discriminant value of $d_{\textrm{B}} > 2.0$, provides an efficiency of $47 \pm 1\%$ for tagging b quarks (as determined from data), and a `mistag' rate of $0.12 \pm 0.01\%$ for misidentifying all lighter (up, down, and strange) quarks and gluons, as b jets. The misidentification rate for charm-quark induced jets is higher than for jets induced by lighter quarks or gluons, and treated separately in the evaluation of systematic uncertainty in the tagging of b jets, with at least two b-tagged jets required in each event.

\subsection{Kinematic Fit}
\label{sec:kinfit}
For the final selection of candidate \ttbar events, a kinematic least-squares ($\chi^{2}$) fit~\cite{blobel,kinfitCMS} is performed to the \ttbar hypothesis. It exploits the characteristic topology of top-quark events, i.e. the presence of two W~bosons that are each reconstructed from the untagged jets and constrained to their accepted mass of \mbox{80.4\GeVcc}~\cite{pdg2012}, and two top quarks reconstructed from the W~bosons and the b-tagged jets. The masses ($m_\cPqt$) of the two top quarks are assumed to be equal, but are not fixed to a specific value so as to use the $m_\cPqt$ distribution to extract an unbiased \ttbar signal, as discussed in Section~\ref{sec:signalbackground}.

To find the most likely combination of six jets, their four-momenta are fitted to the \ttbar final state for each experimentally distinguishable permutation, using all jets in each event that pass the above selection criteria. All b-tagged jets are taken as bottom-quark candidates, and the remaining jets as light-quark candidates. At least two b-tagged and four untagged jets are needed for the fit. For events containing just six jets, two of which are b-tagged, there are six distinguishable jet combinations. When there are more than six jets present in an event, all possible  combinations are considered in the kinematic fitting procedure. If the fit converges for more than one of the possible jet permutations, the one with smallest $\chi^2$ is chosen to represent that event. After the kinematic fit, all events with a fit probability of $P\left(\chi^2\right) > 0.09$ are accepted for further consideration. This cutoff is chosen as it is found to give the smallest combined systematic and statistical uncertainty in simulation.

The kinematic fit to the \ttbar hypothesis assumes Gaussian resolutions for jet energies, pseudorapidities, and azimuthal angles, that are determined separately for jets originating from light and bottom quarks in MC simulated \ttbar events. The resolutions depend on jet energy and pseudorapidity, and are corrected for any differences observed between the data and MC simulation~\cite{CMS-PAS-JME-10-014}.

The number of events remaining in the data after each consecutive selection, and the expected fraction of \ttbar signal, for a \ttbar production cross section of 163\unit{pb}~\cite{Kidonakis:2010dk}, are given in Table~\ref{tab:cutFlow}.

\begin{table}[h]
\begin{center}
\topcaption[Cutflow]{Number of events and the expected fraction of \ttbar events in the data for $\sigma_{\ttbar}=163$\unit{pb}, following each consecutive selection. The expected \ttbar fractions are taken from simulation.}
\begin{tabular}{l r r}
 \\
 \hline
  Selection           & Events   & Fraction of \ttbar \\
 \hline\hline
  At least 6 jets     & 786\,741 & 0.02               \\
  At least two b-tags &  21\,783 & 0.18               \\
  Kinematic fit       &   3\,136 & 0.41               \\
 \hline
\end{tabular}
\label{tab:cutFlow}
\end{center}
\end{table}
\section{Extraction of \texorpdfstring{\ttbar}{ttbar} Signal}
\label{sec:signalbackground}
\begin{figure}[htb!]
\centering
\includegraphics[width=0.49\textwidth]{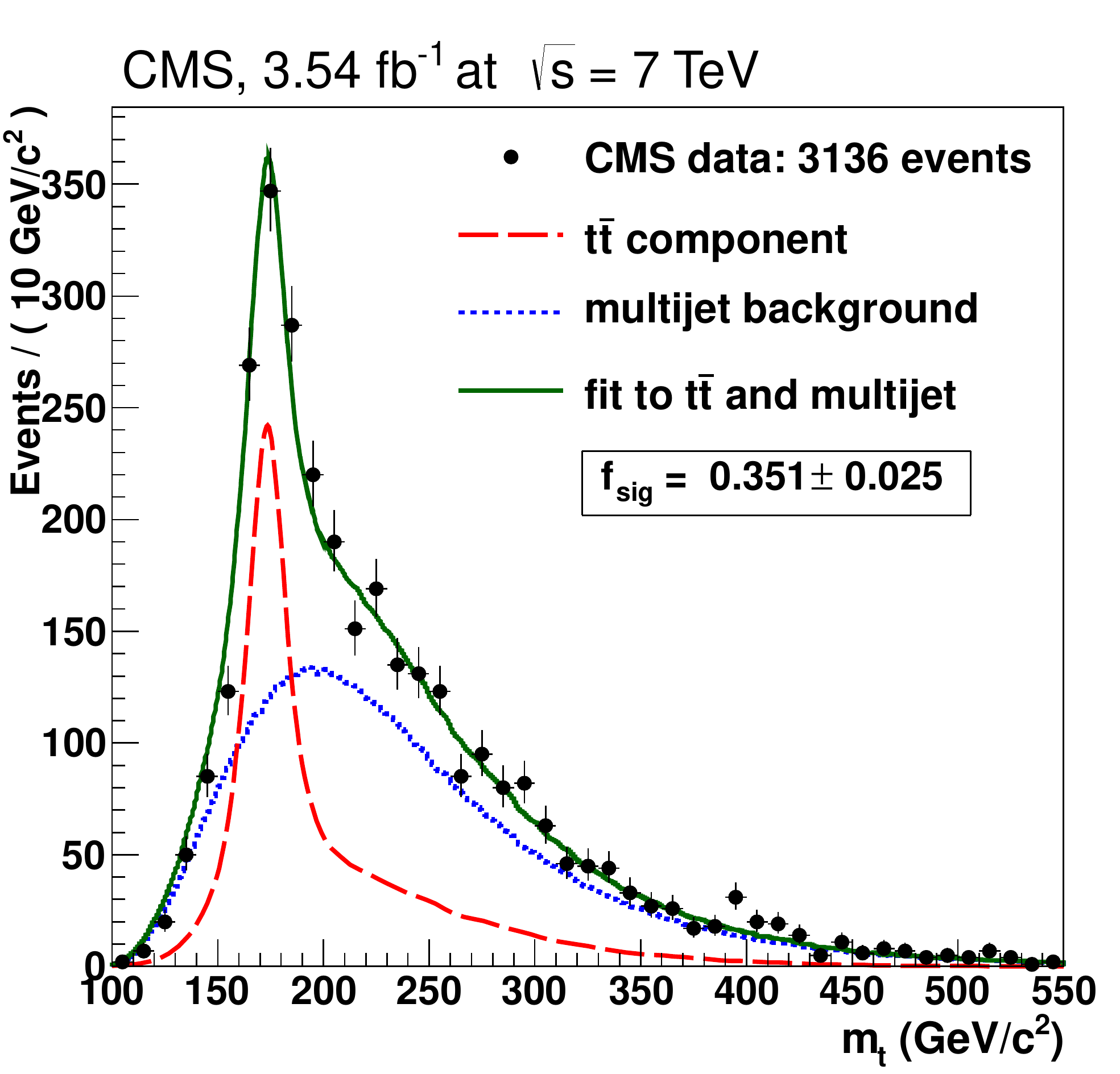}
\caption[Fit Result]{Results of a fit of contributions from a MC \ttbar component (dashed line) and MJ background estimated from data (dotted line) to the distribution of the reconstructed top-quark mass in the data. The uncertainty in the signal fraction represents just the statistical uncertainty obtained from the fit.}
\label{fig:mTopCrossSection}
\end{figure}

The number of \ttbar events remaining after final selections is determined through an unbinned maximum likelihood fit of contributions from \ttbar signal and MJ background (obtained from the MC \ttbar simulation and the multijet data as described in Section~\ref{sec:backgroundestimate}, respectively) to the distribution in reconstructed $m_\cPqt$ obtained from the kinematic fits. The result, shown in Fig.~\ref{fig:mTopCrossSection}, corresponds to a signal fraction ($f_{\text{sig}}$) of $0.351 \pm 0.025$, where the uncertainty is purely statistical. Section~\ref{sec:systematics} provides a discussion of the systematic uncertainties on the MJ background.

\subsection{Estimate of Background from Multijet Events}
\label{sec:backgroundestimate}
The background from multijet production is estimated from data containing ${\geqq}6$ jets using the same criteria as detailed in Section~\ref{sec:eventselection} except for the b-tagging requirement (786\,741 events, as indicated in Table~\ref{tab:cutFlow}). However, as properties of b-tagged and untagged jets differ, those events in this sample that do not have b-tagged jets are weighted to reproduce the distributions appropriate for b-tagged jets from MJ background in the b-tagged \ttbar-candidate data sample. This is done through the jet-tag-rate ratio:
\begin{equation}
  \mathcal{R}(\pt,|\eta|) = \frac{N\left( \pt, |\eta|, d_{\textrm{B}} > 2 \right)}{N\left( \pt, |\eta|, d_{\textrm{B}} < 2 \right)},
\end{equation}
which indicates the relative dependence on jet \pt and jet $|\eta|$ of b-tagged jets and untagged jets in $\geqq 6$-jet MJ events. The numerator reflects the number of b-tagged jets in the MJ data, and the denominator is the number of untagged jets in the same MJ data.
The jet-tag-rate ratio $\mathcal{R}$ is not corrected for contamination of the MJ sample by \ttbar events, as the effect on the signal fraction is below the percent level.

The kinematic fit to the \ttbar hypothesis is then implemented in the ${\geqq}6$-jet MJ events that do not have b-tagged jets, assuming each jet to be a candidate for a b quark, and an event weight is calculated for each fitted permutation of jets with $P\left(\chi^2\right) > 0.09$:
\begin{equation}
  w = \mathcal{R}(\pt^{\cPqb},|\eta^{\cPqb}|) \times \mathcal{R}(\pt^{\cPaqb},|\eta^{\cPaqb}|)
\end{equation}
for the two jets assigned as $\cPqb$ and $\cPaqb$ quarks in the kinematic fit.
This method yields a total of 1\,276\,204 combinations,  which after the weighting with $w$, provides an estimate of the background from MJ events that contain two jets that pass the b-tagging requirement for selection as \ttbar candidates.
This is the distribution for the MJ background shown in Fig.~\ref{fig:mTopCrossSection}. The correction to the distribution due to the application of event weights is typically smaller than the systematic uncertainty assigned to it (see Section~\ref{sec:systematics}).

As the above method is based on properties of single jets, it does not account for any correlations in the background from gluons splitting into $\cPqb$ and $\cPaqb$ quarks. Nevertheless, it appears that these correlations may be negligible as, after subtracting the \ttbar component from the \ttbar candidate sample, there are no significant correlations observed between the fitted top-quark mass for the remaining reconstructed events and the weighted kinematic distributions for the bottom quarks.

\begin{figure}[t]
\centering
\includegraphics[width=0.47\textwidth]{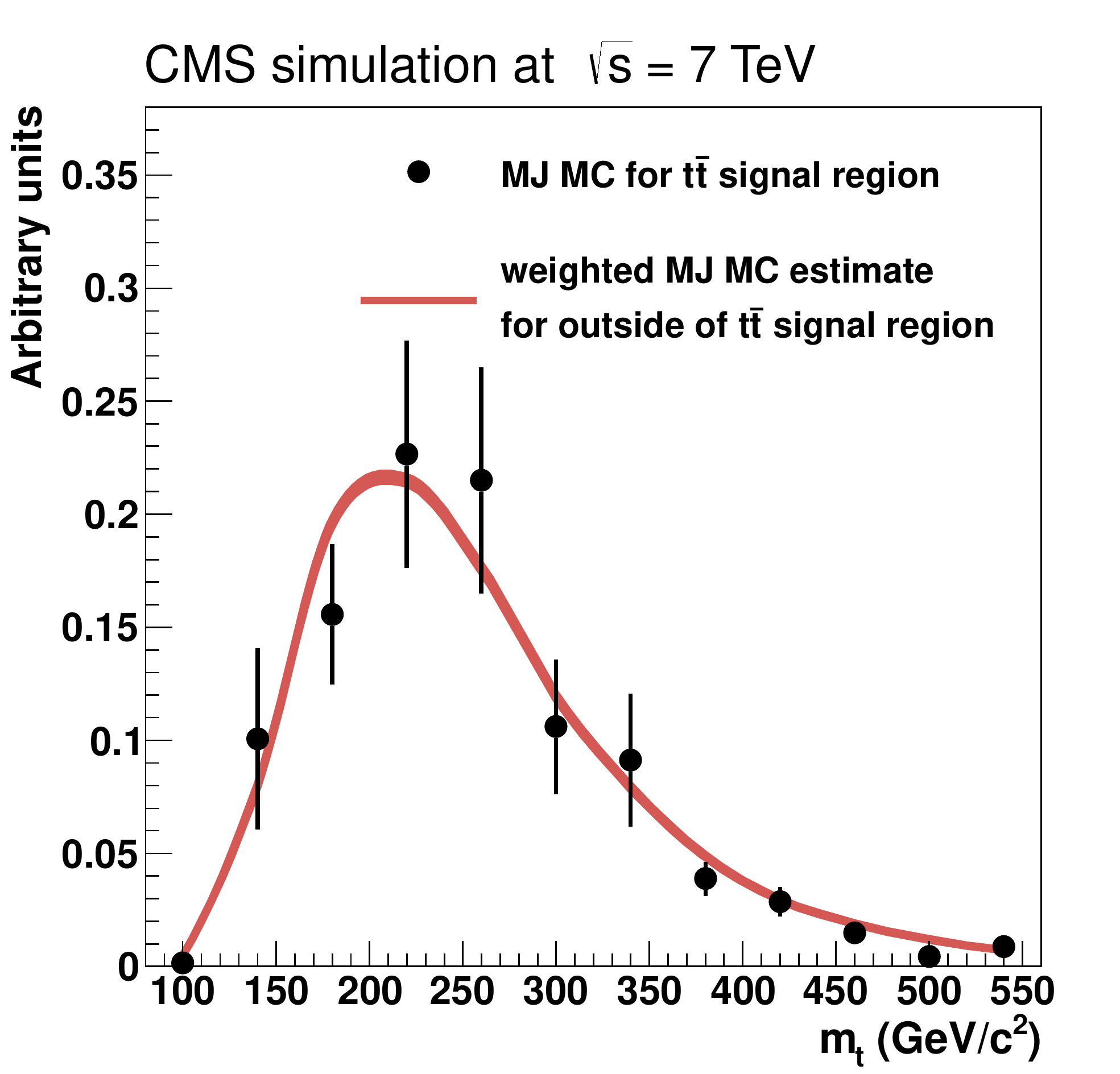}
\includegraphics[width=0.47\textwidth]{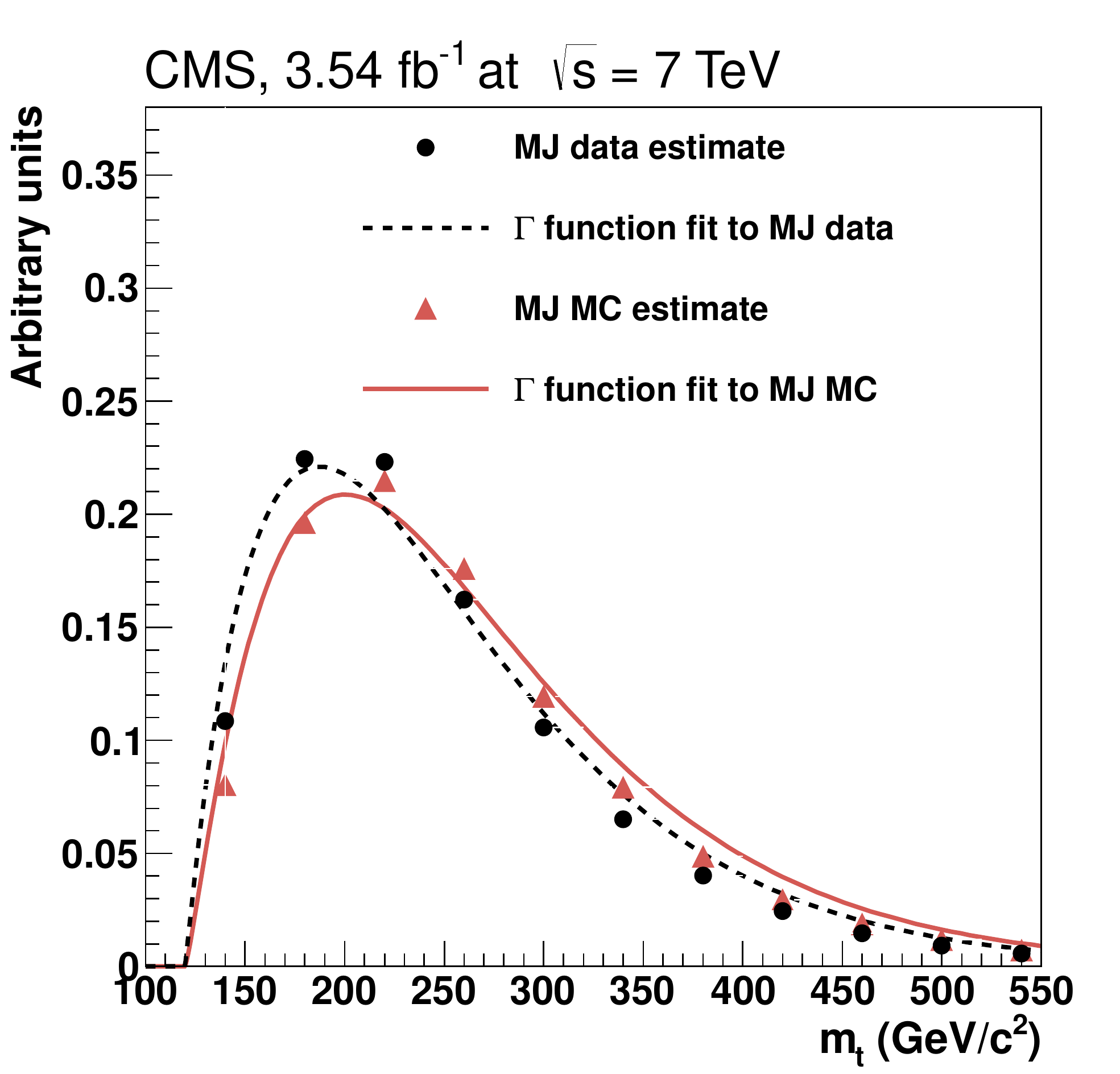}
\caption[Check of Background Estimate]{(left) Comparison of the distribution in reconstructed $m_\cPqt$ for MJ MC events that pass selections for \ttbar candidates (black circles) to weighted events that fail \ttbar selections (thick line). The width of the (narrow) band indicates the 68\% CL of statistical uncertainty. (right) Comparison of the distributions in $m_\cPqt$ for MJ background estimated from the data and from the simulation, with $\Gamma$ functions fitted to each set of points.}
\label{fig:shapeValidation}
\end{figure}

The method is checked by extracting the dependence of $\mathcal{R}(\pt,|\eta|)$ in MC-generated multijet events. Proceeding as before, new weights are applied to simulated events that contain ${\ge}6$ jets, but no b-tagged jets. A comparison of the distributions for $m_\cPqt$ reconstructed in kinematic fits of such MC-weighted events with results of kinematic fits to MJ MC events that pass \ttbar selections is given in Fig.~\ref{fig:shapeValidation}(left). The two distributions are in agreement, thereby confirming the consistency of the method chosen to estimate multijet background in \ttbar candidate events.

\section{Systematic Uncertainties}
\label{sec:systematics}
To determine the systematic uncertainties in the measured cross section, modified samples of simulated events are reanalysed to gauge the impact of changes in the modified parameters. Using MC pseudo-experiments based on the fitted fractions of \ttbar signal and background, we determine the impact on $\sigma_{\ttbar}$ of a change in the value of each parameter (\eg efficiency, signal fraction, or scale of QCD) as described below.

\begin{description}
\item[Jet energy scale:]
The uncertainty from ambiguities in jet energy scale is assessed by shifting the jet energy by ${\pm} 1$~standard deviation (SD) relative to the nominal value, as a function of jet \pt and $\eta$. The method is described in Ref.~\cite{CMS-PAS-JME-10-014}, and is applied with updated values appropriate for these data. The uncertainties in jet energy scale per jet range from 2.0\% to 3.4\%.

\item[Background contribution:]
The systematic uncertainty associated with the normalisation of the background is well within the statistical uncertainty of the fit in Fig.~\ref{fig:mTopCrossSection}. The uncertainty from the distribution of the background as a function of $m_\cPqt$ is determined using a $\Gamma$ function fitted to the background distribution estimated from the MJ data, as follows. The parameters of the $\Gamma$~function are changed by amounts that correspond to the difference observed in the MC simulation of MJ background for the region of \ttbar selection and that estimated using the same technique as used for data in Section~\ref{sec:backgroundestimate}. The uncertainty is defined by half of the difference in the fitted parameters of the $\Gamma$~functions for the background estimates obtained from the MJ data and that from the MC simulation, both shown in Fig.~\ref{fig:shapeValidation}(right).

\item[Tagging of b jets:]
To evaluate the impact of the uncertainty in b tagging on the efficiency of the event selection, the b-tagging efficiency and mistag rate are changed by ${\pm}1$\,SD, corresponding to changes of ${\approx}2\%$ and ${\approx}9\%$ in their values respectively~\cite{CMS-PAPER-BTV-11-004}. When propagated to $\sigma_{\ttbar}$, they change the cross section by ${\pm}6\%$ (cf. Table~\ref{tab:uncertaintyOverview}).

\item[Renormalisation and factorisation scale:]
To study the dependence of the analysis on the renormalisation and factorisation scales ($\mu$) used in the $\ttbar$ MC simulation, the nominal common value $\mu=Q$ for the hard scattering ($Q^2 = m_\cPqt^2 + \sum \pt^2$, where the sum is over all additional final state partons) and for parton showering are simultaneously changed by a factor 0.5 and 2.0. This also reflects the uncertainty in the amount of initial and final-state radiation for changes in the strong coupling constant $\alpha_{s}$ in parton showering by a factor 0.5 and 2.0.

\item[Tune for underlying event:]
The uncertainty in the modeling of the underlying event is based on a comparison of the two \PYTHIA tunes Z2~\cite{tuneZ2} and D6T~\cite{pythiaTunes}, with Z2 as the default tune for this measurement. The uncertainty is defined by half of the difference between the values obtained with these two tunes.

\item[Trigger:]
The trigger efficiency of 96\% observed in the data (Section~\ref{sec:datasets}) is changed by $-$5\% and $+$4\%. These standard deviations are determined by emulating the two triggers used in the subsequent data-taking periods. The energy of the jets reconstructed at the trigger level is also changed by ${\pm} 1$\,SD, which reflects the absolute uncertainty in the jet energy scale for calorimeter jets~\cite{CMS-PAS-JME-10-014}, and provides the ${\pm}1$\,SD limits of uncertainty for trigger efficiencies. Using this procedure, the uncertainties in trigger efficiency are determined in both data and simulation, and are propagated to $\sigma_{\ttbar}$.

\item[Jet energy resolution:]
The jet energy resolution in simulated events is changed by $\pm 10\%$ for $|\eta| < 1.5$, ${\pm}15\%$ for $1.5 < |\eta| < 2.0$, and ${\pm}20\%$ for $|\eta| > 2.0$ The uncertainty is defined by the difference in $\sigma_{\ttbar}$ for the ${\pm}1$\,SD excursions in resolution relative to the nominal values~\citep{CMS-PAS-JME-10-014}.

\item[Matching of parton showers to matrix elements:]
The threshold of the matching scale used for interfacing the matrix elements generated with \MADGRAPH and \PYTHIA parton showering in simulating \ttbar events is changed from the default value of 20\GeV to 10\GeV and to 40\GeV, and propagated to $\sigma_{\ttbar}$.

\item[Mass of the top quark:]
The influence of the value of $m_\cPqt$ is estimated by shifting $m_\cPqt$ in the \ttbar simulation from the nominal 172.5\GeVcc by ${\pm}0.9$\GeVcc, the uncertainty in the currently accepted value of $m_\cPqt$~\cite{TopMassTevatron}.

\item[Pileup:]
The effect of pileup from simultaneous pp interactions is evaluated by superimposing additional minimum-bias events on the simulated signal (on average, ${\approx}8$ observed in the data). To account for uncertainties associated with the measured total inelastic pp cross section, the mean number of observed interactions, and the weighting procedure, the average number of additional pileup events is changed by ${\pm}8\%$, and the impact of the changes extrapolated to $\sigma_{\ttbar}$.

\item[Luminosity:]
The total integrated luminosity is determined with a precision of ${\pm}2.2\%$~\cite{CMS-PAS-SMP-12-008}.
\end{description}

An overview of the different uncertainties contributing to $\sigma_{\ttbar}$ is given in Table~\ref{tab:uncertaintyOverview}. All uncertainties are reasonably symmetric around the mean value of $\sigma_{\ttbar}$. They are therefore averaged and presented as symmetrical excursions about the extracted value of $\sigma_{\ttbar}$. The total uncertainty is obtained by summing the individual uncertainties in quadrature.

\begin{table}[htb]
\begin{center}
\topcaption[List of Uncertainties]{List of all non-negligible uncertainties contributing to the measurement of $\sigma_{\ttbar}$.}
\begin{tabular}{l r}
 \\
 \hline
Source&Relative uncertainty (\%) \\
 \hline
 \hline
  Jet energy scale                        &  10.1 \\
  Background contribution                 &   9.0 \\
  Tagging of b jets                       &   6.0 \\
  Renormalisation and factorisation scale &   5.8 \\
  Tune for underlying event               &   5.5 \\
  Trigger                                 &   5.0 \\
  Jet energy resolution                   &   4.0 \\
  Matching matrix elements/parton showers &   4.0 \\
  Mass of the top quark                   &   2.1 \\
  Pileup                                  &   0.8 \\
 \hline
  Total systematic                        &  18.6 \\
  Total statistical                       &   7.0 \\
  Luminosity                              &   2.2 \\
 \hline
 \hline
 Total uncertainty                        &  20.0 \\
 \hline
\end{tabular}
\label{tab:uncertaintyOverview}
\end{center}
\end{table}
\section{Results}
\label{sec:results}
The $\ttbar$ production cross section, as measured in the all-jet final state, is given by:
\begin{equation}
  \sigma_{\ttbar} = \frac{f_{\text{sig}}\cdot N}{\epsilon \cdot L_{\text{int}}},
\end{equation}
where $f_{\text{sig}}$ is estimated from the fit in Fig.~\ref{fig:mTopCrossSection}. The total number of candidate events is $N=3\,136$, as given in Table~\ref{tab:cutFlow}. The efficiency for selecting \ttbar events as determined from simulation is $\epsilon=0.22\%$ and refers to all possible \ttbar final states. The latter, which is small mainly due to the restrictive jet-\pt selection criteria, includes correction factors for b-tagging efficiency, mistag rate, and trigger efficiency, all obtained from data. The integrated luminosity of the data sample is $L_{\text{int}}= 3.54\fbinv$. These values yield a \ttbar production cross section for an assumed top-quark mass of $m_\cPqt=172.5$\GeVcc of:
\begin{equation}
  \sigma_{\ttbar} = 139 \pm 10\stat \pm 26\syst\pm 3\lum\unit{pb},
\end{equation}
which corresponds to a total uncertainty of $\pm 20\%$, with the individual contributions listed in Table~\ref{tab:uncertaintyOverview}.
\section{Alternative Analysis using a Neural-Network-Based Selection}
\label{sec:crosscheck}

A separate measurement of $\sigma_{\ttbar}$ is also performed as a cross-check. The kinematic properties of
signal and background events are used to develop selection criteria based on a neural-network (NN) procedure,
which is expected to be less sensitive to the jet energy scale (JES).
In addition, the MJ background is estimated using a model that takes account of
correlations between jets that pass b-tagging criteria.

\begin{figure}[b]
\includegraphics[width=.47\textwidth]{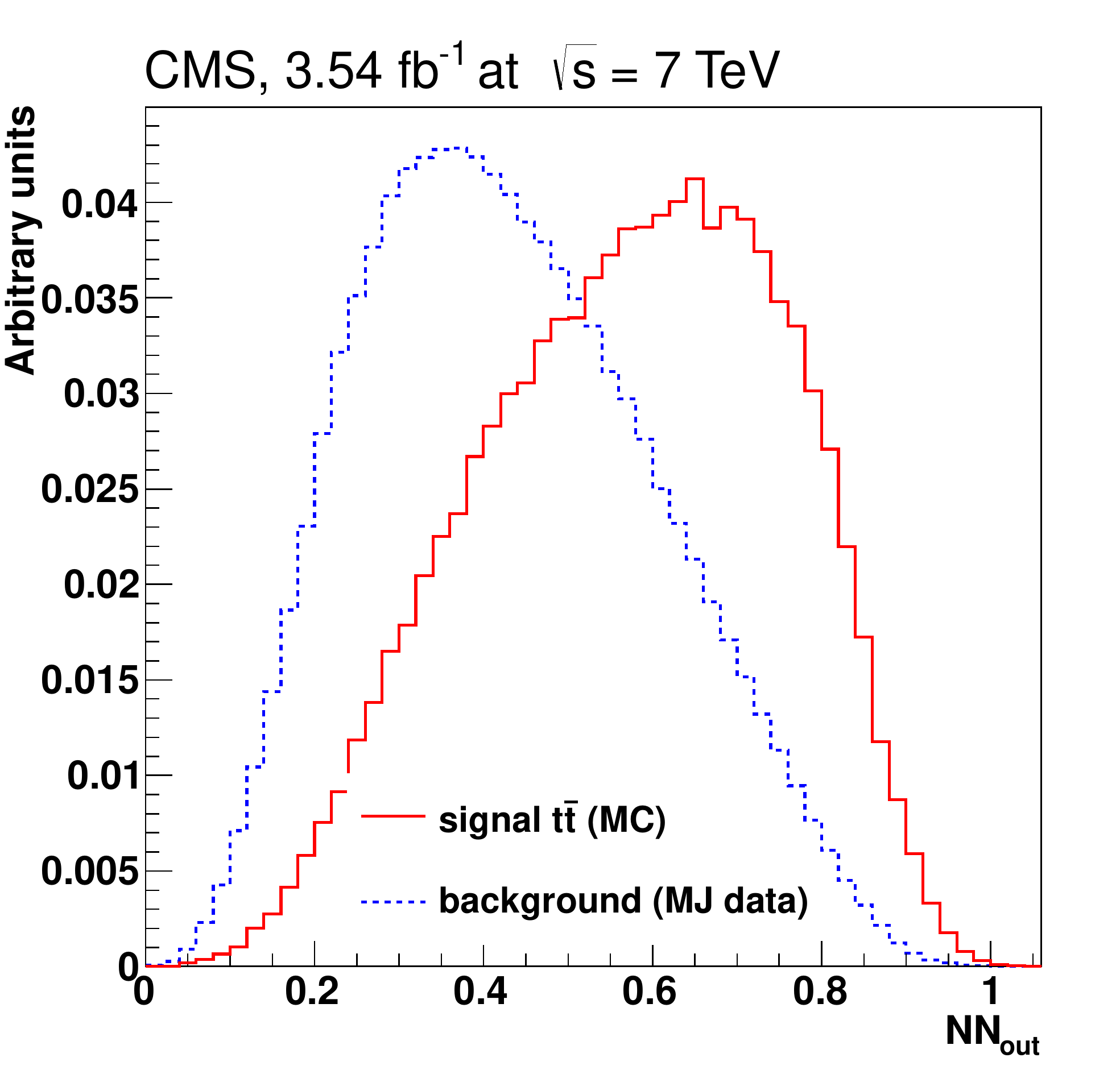}
\includegraphics[width=.47\textwidth]{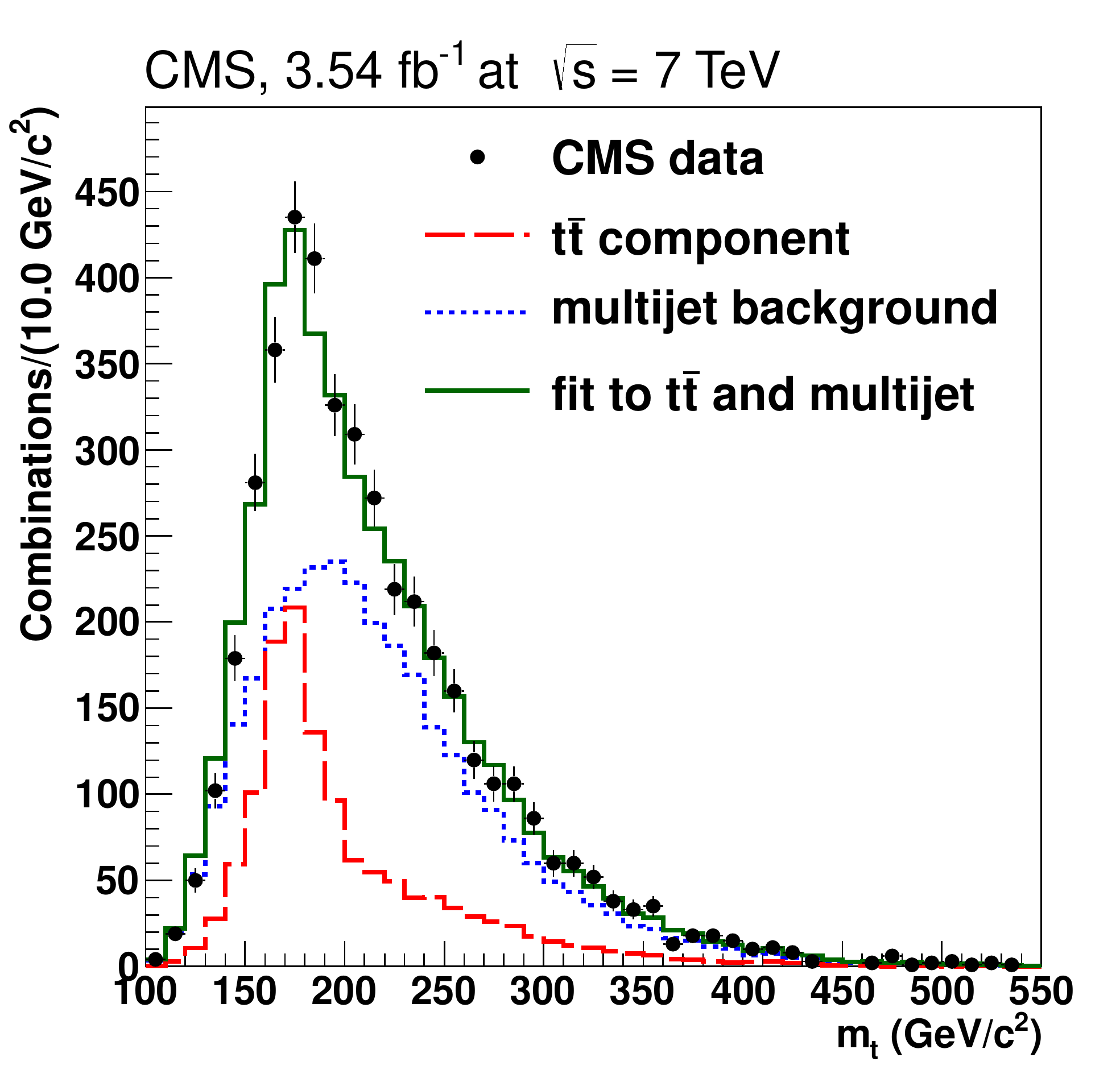}
\caption{(left) Neural network output for the \ttbar MC and multijet background components shown normalised to
unity. (right) Result of a fit of contributions from a MC \ttbar component (dashed line) and MJ background
estimated from data (dotted line) to the distribution of the reconstructed top-quark mass in the data.}
\label{fig:crosscheck}
\end{figure}

The NN is trained on a set of simulated \ttbar events with jet multiplicities
of
$6 \leqq N_{\textrm{jet}} \leqq 8$ (using the same criteria of
 Section~\ref{sec:eventselection}) and an equal number of MJ events of same range
in jet multiplicity that have a greatly reduced \ttbar component relative to background.
Six variables are used to train the neural network. One, called centrality, is defined as the
ratio of the scalar sum of the transverse energies ($\et = E \sin \theta$) of the jets
to the invariant mass of the multijet system (MJS). Another variable is the aplanarity, defined
as $\frac{3}{2}\mathcal{Q}_{1}$, where $\mathcal{Q}_{1}$ is the smallest of the three normalised
eigenvalues of the sphericity tensor $M_{ab}=\sum_j p^{a}_jp^{b}_j$,
calculated in the centre-of-mass of the MJS, where $a$ and $b$ reflect the three
spatial components of the momentum of each jet $p_j$. The remaining variables are defined by the
ratio of \et values of the two jets of largest \et to the scalar sum of the transverse energies of
all jets, the transverse energies of these jets
multiplied by $\sin^2 \theta^*$ ($\theta^*$ being the angle between the jet and the beam axis
in the centre-of-mass of the MJS), and the average of the scalar
quantity $\et \sin^2\theta^*$ for the remaining jets.
The output of the neural network
 ($\mathrm{NN}_{\text{out}}$),  shown in Fig.~\ref{fig:crosscheck}(a), is used to enhance the
\ttbar signal purity by requiring
$\mathrm{NN}_{\text{out}} > 0.65$. To compensate for this restrictive selection, a more
efficient b-tagging algorithm is used \cite{CMS-PAPER-BTV-11-004} in this analysis, requiring
the b-jet discriminant $d^\prime_\mathrm{B}$ for the track in the jet with an
impact parameter of next-to-highest significance to be $d^\prime_\mathrm{B} > 3.3$, which improves
by about 30\% the b-tagging efficiency. Candidate events are required to have at least
two such
b-tagged jets.

The multijet background estimate is obtained from the data, inferring it from a sample of events with just
five jets. The probability of any two jets in the background sample to pass the b-tag condition for the
 \ttbar candidate selection is defined relative to pairs of jets that both pass a looser b-tag requirement
 ($d^\prime_\mathrm{B} > 1.7$). This probability is defined by the ratio
\begin{equation}
  \mathcal{R}_{\mathrm{LL}}^{\mathrm{MM}}(\langle \pt\rangle, ~\langle |\eta| \rangle,~ \Delta R) = \frac{N(\langle \pt\rangle, ~\langle |\eta| \rangle,~ \Delta R,~ d^\prime_{\mathrm{B}} > 3.3)}{N(\langle \pt\rangle, ~\langle |\eta| \rangle,~ \Delta R,~ d^\prime_{\mathrm{B}} > 1.7)},
\end{equation}
parameterised in terms of the average transverse momentum $\langle \pt \rangle$ of the two jets,
the average of the absolute values of the two pseudorapidities $\langle |\eta| \rangle$, and
the $\Delta R = \sqrt{(\Delta\eta)^{2}+(\Delta\varphi)^{2}}$ separation between the two jets. The expected
background from pairs of b-tagged jets is obtained by
weighting each pair of less restrictive b-tags with the corresponding $\mathcal{R}_{\mathrm{LL}}^{\mathrm{MM}}$.

The top-quark mass is reconstructed from a kinematic fit, with minor modifications with respect to the one
described in Section~\ref{sec:kinfit}.
All permutations where two b-tagged jets are associated to the b quarks are considered.
To increase \ttbar purity and reduce false permutations, a cut is made on the quality of the fit and on
 the separation between the two b-tagged jets, $\Delta R > 1.5$.
Following event selection, 3\,571~events remain in the sample, corresponding to 4\,329 fitted $m_\cPqt$
values. The expectation for $m_\cPqt$ in \ttbar events is acquired directly from simulation,
while the MJ background is obtained from data, as described above, but correcting for the presence
of \ttbar events in the background sample. Finally, the cross section is extracted from a binned maximum
likelihood fit of these two contributions to the data, see Fig.~\ref{fig:crosscheck}(b).
The measured cross section is
$\sigma_{\ttbar}= 114 \pm 15\stat \pm 27\syst \pm 3\lum\unit{pb}$, with a
signal fraction of $30 \pm 4\%$.

The obtained value for the cross section is well within 2\,SD of that measured in the main analysis.
Considering the difference in event selections, the two measurements can be regarded as compatible.
However, the results are also partly independent, which implies that they could be combined to improve the uncertainty on the cross section.
For instance, the uncertainty associated with JES is about 40\% smaller for the alternative method.
However the statistical uncertainty on the second result is larger, which reflects the fact that this analysis has more background and less difference in the distributions of $m_\cPqt$ for signal and background events.
Because of that poorer statistical significance, the second result is not combined with that from the main analysis.

\section{Summary}
\label{sec:summary}

Assuming a top-quark mass of 172.5\GeVcc, a first measurement of the $\ttbar$ production cross section in the all-jet channel at $\sqrt{s} = 7\TeV$ yields $\sigma_{\ttbar} = 139 \pm 10\stat \pm 26\syst \pm 3\lum\unit{pb}$.

This result is consistent within 2\,SD of an alternative analysis using a neural-network-based selection, and with previous CMS measurements in dilepton and lepton+jets final states, as well as with the predictions of the standard model. The most precise single CMS measurement is currently in the dilepton channel and provides $\sigma_{\ttbar}=161.9\pm2.5\stat {}^{+5.1}_{-5.0}\syst\pm3.6\lum\unit{pb}$~\cite{top-11-005}. Two predictions of the SM based on approximate next-to-next-to-leading-order calculations yield $\sigma_{\ttbar}=164^{+10}_{-13}$\unit{pb}~\cite{Aliev:2010zk} and $\sigma_{\ttbar}=163^{+11}_{-10}$~pb~\cite{Kidonakis:2010dk}.

This measurement complements the set of CMS measurements of \ttbar production at the LHC.
\section*{Acknowledgements}
We congratulate our colleagues in the CERN accelerator departments for the excellent performance of the LHC and thank the technical and administrative staffs at CERN and at other CMS institutes for their contributions to the success of the CMS effort. In addition, we gratefully acknowledge the computing centres and personnel of the Worldwide LHC Computing Grid for delivering so effectively the computing infrastructure essential to our analyses. Finally, we acknowledge the enduring support for the construction and operation of the LHC and the CMS detector provided by the following funding agencies: BMWF and FWF (Austria); FNRS and FWO (Belgium); CNPq, CAPES, FAPERJ, and FAPESP (Brazil); MEYS (Bulgaria); CERN; CAS, MoST, and NSFC (China); COLCIENCIAS (Colombia); MSES (Croatia); RPF (Cyprus); MoER, SF0690030s09 and ERDF (Estonia); Academy of Finland, MEC, and HIP (Finland); CEA and CNRS/IN2P3 (France); BMBF, DFG, and HGF (Germany); GSRT (Greece); OTKA and NKTH (Hungary); DAE and DST (India); IPM (Iran); SFI (Ireland); INFN (Italy); NRF and WCU (Korea); LAS (Lithuania); CINVESTAV, CONACYT, SEP, and UASLP-FAI (Mexico); MSI (New Zealand); PAEC (Pakistan); MSHE and NSC (Poland); FCT (Portugal); JINR (Armenia, Belarus, Georgia, Ukraine, Uzbekistan); MON, RosAtom, RAS and RFBR (Russia); MSTD (Serbia); SEIDI and CPAN (Spain); Swiss Funding Agencies (Switzerland); NSC (Taipei); ThEP, IPST and NECTEC (Thailand); TUBITAK and TAEK (Turkey); NASU (Ukraine); STFC (United Kingdom); DOE and NSF (USA).
Individuals have received support from the Marie-Curie programme and the European Research Council (European Union); the Leventis Foundation; the A. P. Sloan Foundation; the Alexander von Humboldt Foundation; the Belgian Federal Science Policy Office; the Fonds pour la Formation \`a la Recherche dans l'Industrie et dans l'Agriculture (FRIA-Belgium); the Agentschap voor Innovatie door Wetenschap en Technologie (IWT-Belgium); the Ministry of Education, Youth and Sports (MEYS) of Czech Republic; the Council of Science and Industrial Research, India; the Compagnia di San Paolo (Torino); and the HOMING PLUS programme of Foundation for Polish Science, cofinanced from European Union, Regional Development Fund.

\bibliography{auto_generated}   

\begin{thebibliography}{10}
\bibitem{tuneZ2}
\hrefCMSnoop {} {}The {\sc pythia6} {\sc Z2} tune is identical to the {\sc Z1}
  tune described in~\cite{pythiaTunes} except that {\sc Z2} uses the CTEQ6L1
  parton distribution functions while {\sc Z1} uses CTEQ5L.
\end{thebibliography}

\cleardoublepage \appendix\section{The CMS Collaboration \label{app:collab}}\begin{sloppypar}\hyphenpenalty=5000\widowpenalty=500\clubpenalty=5000\textbf{Yerevan Physics Institute,  Yerevan,  Armenia}\\*[0pt]
S.~Chatrchyan, V.~Khachatryan, A.M.~Sirunyan, A.~Tumasyan
\vskip\cmsinstskip
\textbf{Institut f\"{u}r Hochenergiephysik der OeAW,  Wien,  Austria}\\*[0pt]
W.~Adam, E.~Aguilo, T.~Bergauer, M.~Dragicevic, J.~Er\"{o}, C.~Fabjan\cmsAuthorMark{1}, M.~Friedl, R.~Fr\"{u}hwirth\cmsAuthorMark{1}, V.M.~Ghete, J.~Hammer, N.~H\"{o}rmann, J.~Hrubec, M.~Jeitler\cmsAuthorMark{1}, W.~Kiesenhofer, V.~Kn\"{u}nz, M.~Krammer\cmsAuthorMark{1}, I.~Kr\"{a}tschmer, D.~Liko, I.~Mikulec, M.~Pernicka$^{\textrm{\dag}}$, B.~Rahbaran, C.~Rohringer, H.~Rohringer, R.~Sch\"{o}fbeck, J.~Strauss, A.~Taurok, W.~Waltenberger, G.~Walzel, E.~Widl, C.-E.~Wulz\cmsAuthorMark{1}
% ...
\vskip\cmsinstskip
\dag:~Deceased\\
1:~~Also at Vienna University of Technology, Vienna, Austria\\
% ...

\end{sloppypar}
\end{document}